\documentclass[fleqn,usenatbib]{mnras}
\pdfoutput=1

\usepackage{amssymb}
\usepackage{amsmath}

\usepackage{graphicx}
\usepackage{graphics}
\usepackage{dcolumn}
\usepackage[dvipsnames]{xcolor}
\usepackage{fancyhdr} 
\usepackage{graphicx}
\usepackage{float}
\usepackage{multirow}

\usepackage{subfig}
\usepackage{hyperref}

\usepackage[utf8]{inputenc}

\usepackage{dsfont}

\usepackage[justification=centerlast, format=plain, labelfont=bf]{caption}

\makeatletter
\renewcommand\tableofcontents{%
    \@starttoc{toc}%
}
\makeatother

\DeclareCaptionJustification{justified}{\leftskip=0pt \rightskip=0pt \parfillskip=0pt plus 1fil}

\def\VEV#1{\left\langle #1 \right\rangle}

    \newcommand{\be}{\begin{equation}}
  \newcommand{\ee}{\end{equation}}
    \newcommand{\ba}{\begin{align}}
  \newcommand{\ea}{\end{align}}

\newcommand{\MUV}{ M_{\rm UV} }
\newcommand{\fesc}{ f_{\rm esc} }
\newcommand{\xHI}{ x_{\rm HI} }
\newcommand{\xHII}{ x_{\rm HII} }
\newcommand{\nH}{ n_{\rm H} }

\usepackage{newtxtext,newtxmath}

\usepackage[T1]{fontenc}

\DeclareRobustCommand{\VAN}[3]{#2}
\let\VANthebibliography\thebibliography
\def\thebibliography{\DeclareRobustCommand{\VAN}[3]{##3}\VANthebibliography}

\usepackage{graphicx}	
\usepackage{amsmath}	

\title[A JWST photon budget crisis?]
{Reionization after JWST: a photon budget crisis?}

\author[J.~B.~Muñoz et al.]{
Julian B.~Muñoz$^{1}$\thanks{E-mail: julianbmunoz@utexas.edu},
Jordan Mirocha$^{2,3}$,
John Chisholm$^{1}$,
Steven R. Furlanetto$^{4}$,
and Charlotte Mason$^{5}$
\\
$^{1}$Department of Astronomy, The University of Texas at Austin, 2515 Speedway, Stop C1400, Austin, TX 78712, USA \\
$^{2}$Jet Propulsion Laboratory, California Institute of Technology, 4800 Oak Grove Drive, Pasadena, CA 91109, USA \\
$^{3}$California Institute of Technology, 1200 E. California Boulevard, Pasadena, CA 91125, USA \\
$^{4}$Department of Physics and Astronomy, University of California, Los Angeles, CA 90095, USA \\
$^{5}$Niels Bohr Institute, University of Copenhagen, Jagtvej 128, 2200 København N, Denmark \\
}

\pubyear{-}

\begin{document}
\label{firstpage}
\pagerange{\pageref{firstpage}--\pageref{lastpage}}
\maketitle

\begin{abstract}
New {\it James Webb} Space Telescope (JWST) observations are revealing the first galaxies to be prolific producers of ionizing photons, which we argue gives rise to a tension between different probes of reionization.
Over the last two decades a consensus has emerged where star-forming galaxies are able to generate enough photons to drive reionization, given reasonable values for their number densities, ionizing efficiencies $\xi_{\rm ion}$ (per unit UV luminosity), and escape fractions $f_{\rm esc}$.
However, some new JWST observations infer high values of $\xi_{\rm ion}$ during reionization and an enhanced abundance of earlier ($z\gtrsim 9$) galaxies, 
dramatically increasing the number of ionizing photons produced at high $z$.
Simultaneously, recent low-$z$ studies predict significant escape fractions for faint reionization-era galaxies.
Put together, we show that the galaxies we have directly observed ($M_{\rm UV} < -15$) not only can drive reionization, but would end it too early.
That is, our current galaxy observations, taken at face value, imply an excess of ionizing photons and thus a process of reionization in tension with the cosmic microwave background (CMB) and Lyman-$\alpha$ forest.  
Considering galaxies down to $M_{\rm UV}\approx -11$, below current observational limits, only worsens this tension.
We discuss possible avenues to resolve this photon budget crisis, including systematics in either theory or observations.
\end{abstract}




\section{Introduction}

The epoch of reionization represents the last major phase transition of our universe.
During reionization the intergalactic gas went from cold and neutral before the first cosmic structures formed (at redshift $z\sim 30$, or 100 Myrs after the Big Bang) to hot and ionized by $z\sim 5$ (roughly a billion years later).
While we are certain that this process took place, we do not know how. 
The likely culprits for reionization are the first star-forming galaxies \defcitealias{Robertson:2015uda}{R15}\citep[][hereafter \citetalias{Robertson:2015uda}]{Robertson:2015uda}, but other suspects include supermassive black holes~\citep{Madau:2015cga,Madau:2024fzo}, and even dark matter~\citep{Liu:2016cnk}.
More broadly, the timing and topology of reionization hold a treasure trove of information on the astrophysics of the early universe, which we have yet to uncover.

The accounting of reionization is rather simple: there have to be enough photons to ionize all the intergalactic hydrogen atoms, including their recombinations.
During the WMAP era this was a stringent requirement, as cosmic microwave background (CMB) data implied an approximate midpoint of reionization at $z=10-11$~\citep{WMAP:2010qai}, earlier than expected from standard galaxy-formation models and beyond the reach of contemporaneous direct observations.
With the advent of the {\it Planck} satellite  this tension was eased, as newer CMB data preferred later reionization (with an effective $z\sim 7-8$,~\citealt{Planck:2015fie}), and by then {\it Hubble} Space Telescope (HST) observations had characterized a population of star-forming galaxies at those redshifts~\citep{Madau:2014bja}. Together, these observations alleviated the demand for ionizing photons and quickly led to the consensus that, under reasonable assumptions, star-forming galaxies were able to drive reionization~(\citetalias{Robertson:2015uda}, \citealt{Bouwens:2015vha},\defcitealias{Finkelstein:2019sbd}{F19}\citealt[][hereafter \citetalias{Finkelstein:2019sbd}]{Finkelstein:2019sbd}).
In this {\it Letter} we examine whether this consensus holds in light of recent {\it James Webb} Space Telescope (JWST) observations of the high-redshift universe.

Three key factors determine the average reionization history: the production rate of ionizing photons (by early galaxies and black holes), the fraction $f_{\rm esc}$ of those photons that escape to the intergalactic medium (IGM) and can ionize neutral hydrogen, and the number of recombinations per hydrogen atom.
While there remain open questions about the last factor~\citep{Davies:2021inu}, the first two are particularly uncertain.

The production rate of ionizing photons is given by the early-galaxy abundance, usually expressed through the UV luminosity function (UVLF, the comoving number density of galaxies per UV magnitude), times the ionizing efficiency $\xi_{\rm ion}$ of each galaxy.
Though there is broad agreement on the bright end of the UVLF, the number density of ultra-faint (below $\MUV \approx -14$) galaxies is virtually unconstrained.
Theoretically, we expect the UVLF to ``turn over'' at some magnitude $\MUV^{\rm turn}$ due to feedback~\citep{Shapiro:2003gxa}, and HST observations have constrained this turnover to be fainter than $\MUV^{\rm turn}\approx-15$~\citep{Atek:2018nsc}.
At the same time, some new JWST observations are finding early galaxies to have higher ionizing efficiencies $\xi_{\rm ion}$ than canonically assumed [with $\log_{10}\xi_{\rm ion}/$(Hz erg$^{-1}$) $\approx 25.5-26.0$ vs 25.2 \citealt{Atek24_Nature_xiion,Simmonds24_xiion,Endsley23_reionization,Prieto24_xiion,CurtisLakeJades_xiion,Hsiao23_xiion,Calabro24_glass_xiion}, though see~\citealt{Matthee:2022rbw,Meyer24,Pahl:2024utu}].
Moreover, JWST is also unveiling an enhanced population of both star-forming galaxies at $z\gtrsim 9$~\citep[][with an unknown origin~\citealt{Mason:2022tiy,Ferrara2022,Munoz:2023cup,Mirocha_UVLFs2023}]{Finkelstein_CEERS,Finkelstein_CEERS2,Eisenstein_JADES,Harikane_UVLFs,Castellano_GLASS_hiz} and supermassive black holes~\citep[][though they are likely obscured \citealt{Greene24_AGN}]{Matthee:2023utn}, which would further boost the ionizing-photon budget.
Such a wealth of photons will accelerate the process of reionization, if they escape their host galaxies.

The escape fraction $\fesc$ of early galaxies is a contentious topic.
The basic problem is that the ionizing-photon production is dominated by very massive, short-lived stars, which may live and die before their birth clouds are dispersed, minimizing photon escape. 
The BPASS models \citep{Eldridge2009} provided a new hope for high escape fractions, as binary interactions help to lengthen effective stellar lifetime and so boost 
the effective $\fesc$. 
Models in which the escape fraction is set by local, cloud-scale physics, suggest that $\fesc$ could be independent of galaxy properties like mass or luminosity \citep{Ma2016}. 
However, different simulations predict $f_{\rm{esc}}$ growing for brighter galaxies~\citep{Sharma2016}, declining~\citep{Wise2014,Kimm2014}, or peaking at intermediate masses~\citep[e.g.,][]{Yoo2020,Ma2020,Rosdahl2022,Yeh2023}.
From a theoretical perspective, there seems to be no clear consensus on the nature of $f_{\rm{esc}}$ in high-$z$ galaxies.
Observationally, it is extremely challenging to measure $\fesc$ while there is neutral hydrogen in the IGM.
However, recent studies of low-$z$ analogues of reionization-era galaxies have found a strong correlation between their escape fractions and UV slopes $\beta_{\rm UV}$: bluer galaxies exhibit larger values of $f_{\rm esc}$~\citep{Flury:2022asy,Chisholm22_betafesc,Begley22_VANDELSfesc,Saldana-Lopez:2022niz}.
JWST and HST data show that early galaxies have bluer slopes than their average low-$z$ counterparts~\citep[e.g.,][]{Topping23_jwst_UVSlopes,Cullen23_jwst_UVSlopes,Weibel24_betaUV}, such that the few studies of reionization-era galaxies indicate modest $\fesc$ values near 5--15\% \citep{mascia23, lin24}.

Here we argue that combining the abundance of directly observed reionization-era galaxies, the new JWST estimates of $\xi_{\rm ion}$, and the low-$z$ insights on $\fesc$ leads to too many ionizing photons at high redshifts, ending reionization too early.
Such an early reionization is in contradiction with current CMB~\citep[][]{Planck:2018vyg} and Lyman-$\alpha$ forest observations~\citep{Bosman:2021oom}, 
and poses {\it a tension in the photon budget during reionization}.
We will outline possible ways to ease this tension, including physical ingredients missing in our theoretical models, interpretation of observations, or both.

Through this paper we assume a flat $\Lambda$CDM cosmology with $h=0.7$ and $\Omega_M=0.3$ to match that assumed in~\citet{Bouwens21} and~\citet{Donnan24_UVLF}, 
all magnitudes are AB~\citep{OkeGunnAB}, and quantities are spatially averaged unless otherwise indicated.

\defcitealias{Matthee:2021qmg}{M22}

\begin{figure}
    \centering
    \includegraphics[width=\linewidth]{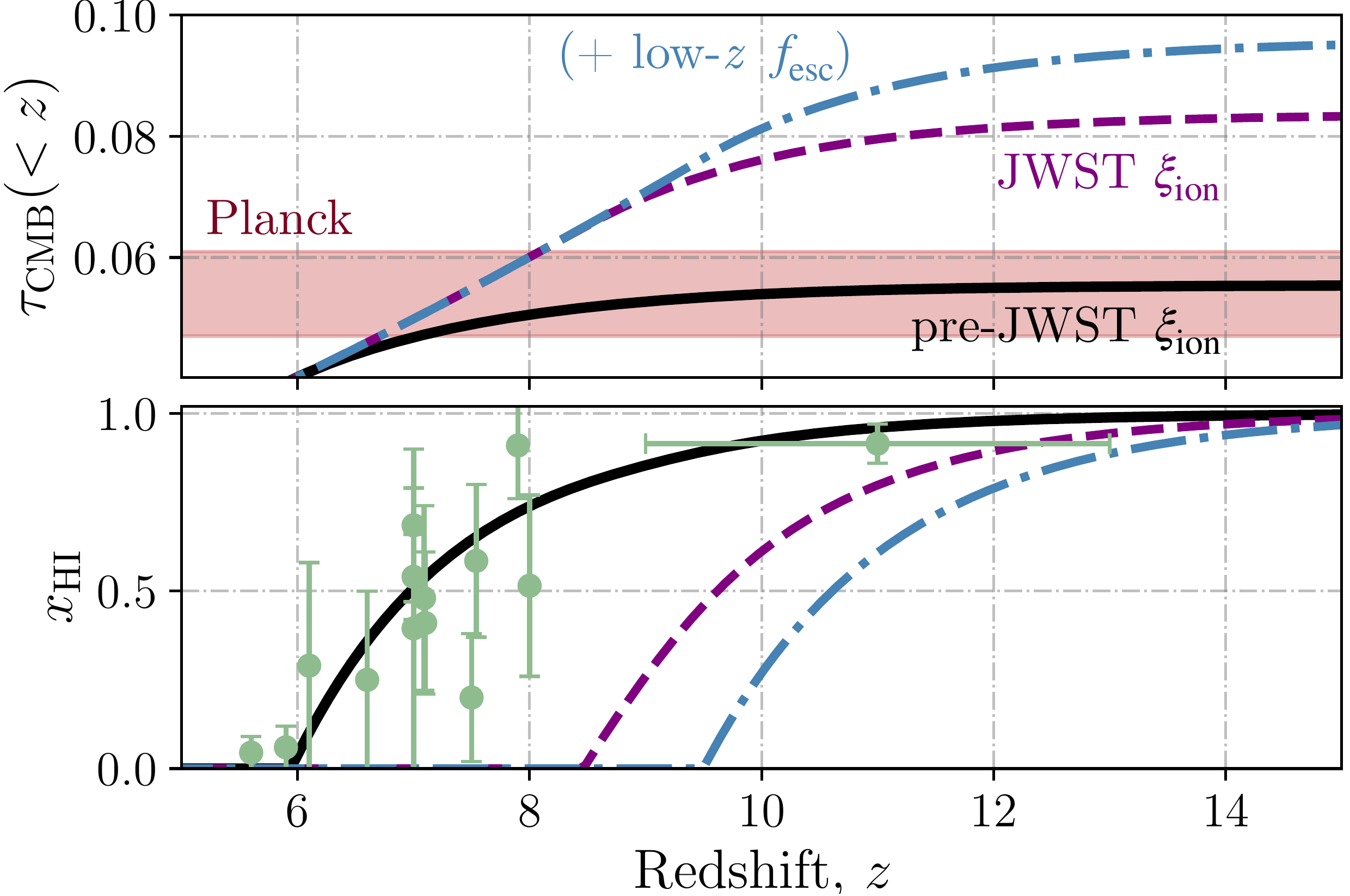}
    \caption{The new JWST and low-$z$ observations imply an earlier reionization, in tension with the CMB. {\bf Bottom:} Evolution of the neutral fraction $\xHI$ as a function of redshift $z$ for a pre-JWST model (black solid, with a cutoff at $\MUV=-13$ and $\fesc=0.2$, following \citetalias{Robertson:2015uda}), 
    for the same model but with a JWST-calibrated $\xi_{\rm ion}$ (purple dashed, following~\citealt{Simmonds24_xiion}), and a model where in addition $\fesc$ is determined from low-$z$ analogues (blue dot-dashed, using the fit in~\citealt{Chisholm22_betafesc}).
    Green points show a collection of observational constraints from~\citep{McGreer:2014qwa,Greig:2016vpu,Greig:2018rts,Sobacchi:2015gpa,Mason:2019ixe,Whitler:2019nul,Wang:2020zae,Nakane23_xHILAEs_JWST} (see also \citealt{Bruton24_xHI088})
    {\bf Top:} CMB optical depth $\tau_{\rm CMB}$, where the red band is the measurement from~\citet{Planck:2018vyg}.
    The new galaxy observations give rise to far more ionizing photons, and at face value are in severe tension with CMB data.
    }
    \label{fig:xHIz}
\end{figure}

\section{Modeling Reionization}

We will follow a simple model of reionization to solve for the volume-averaged hydrogen neutral fraction $\xHI\equiv n_{\rm HI}/n_{\rm H}$, and its complement the ionized fraction $\xHII \equiv 1 - \xHI$.
This quantity evolves as~\citep{Madau:1998cd}
\be
\dot x_{\rm HII} = \dfrac{\dot n_{\rm ion}}{\nH} - \dfrac{\xHII}{t_{\rm rec}},
\ee
which showcases the competition between the ``sources'' (first term) and ``sinks'' (second) of ionizing photons.
The former is given by the density of ionizing photons produced ($\dot n_{\rm ion}$) divided by that of hydrogen $\nH = \rho_{\rm b}(1-Y_{\rm He})/m_{\rm H}$, where $Y_{\rm He}$ is the Helium mass fraction, $m_{\rm H}$ the proton mass, and $\rho_{\rm b}$ the baryon energy density, which scales as $(1+z)^3$.
The sink term captures the number of recombinations that hydrogen atoms suffer on average, characterized by a timescale~\citep{Shull2012_EoR}
\be
t_{\rm rec} = \left [C~\alpha_{\rm B} (1 + x_{\rm He}) \nH \right]^{-1}
\ee
where $x_{\rm He} \equiv n_{\rm He}/\nH \approx Y_{\rm He}/[4(1-Y_{\rm He})]$, $\alpha_{\rm B}$ is the case-B recombination coefficient, and $C$ is the clumping factor. 
The clumping factor is difficult to estimate theoretically, as it depends on how ionized regions penetrate into high-density clumps. 
Simulations predict $C\approx 2-5$ during reionization, growing towards lower $z$~\citep[e.g.,][]{Pawlik:2015bja}.
Recent work in \citet{Davies:2021inu} instead suggests that more recombinations are needed to explain the short mean free path of ionizing photons at $z \sim 6$ (thanks to absorption by pervasive high-density clumps known as Lyman-limit systems, \citealt{Becker2021, Zhu2023}).
For simplicity and comparison with past literature~(\citetalias{Robertson:2015uda}), we will set $C=3$ and evaluate $\alpha_{\rm B}$ at $T=2\times 10^4$ K for now (which yields nearly identical results to using the $C(z)$ fit from \citealt{Shull2012_EoR}), and 
return to the effect of recombinations later.

For reionization to progress the sources have to win over the sinks.
Our sources will be star-forming galaxies, which produce a background of ionizing photons at a rate of
\be
\dot n_{\rm ion} = \int d\MUV \Phi_{\rm UV} \dot N_{\rm ion} \fesc,
\label{eq:niondot}
\ee
where all factors inside the integral are assumed to depend on $\MUV$, and we integrate down to a cutoff magnitude $\MUV^{\rm ion.\, cutoff}$ that will be a free parameter.
Here, $\Phi_{\rm UV}$ is the UVLF, taken at $z\leq 9$ from the pre-JWST fit in \citet{Bouwens21} and at $z>9$ from the JWST calibrations of \citet[][see Appendix \ref{app:otherUVLF} for alternative analyses using only pre-JWST UVLFs, including that of \citealt{FinkelsteinBagley22_UVLF}]{Donnan24_UVLF}, $\dot N_{\rm ion} \equiv L_{\rm UV} \, \xi_{\rm ion}$
is the production rate of ionizing photons per galaxy, given by their UV luminosity $L_{\rm UV}$ times the ionizing efficiency $\xi_{\rm ion}$, of which a fraction $\fesc$ escapes into the IGM.

It is apparent that the product $\xi_{\rm ion} \times \fesc$ will determine the timing of reionization, and that these two factors are, at face value, fully degenerate.
Increasing $\xi_{\rm ion}$ per galaxy while decreasing $f_{\rm esc}$ will yield identical effects on the IGM.
Fortunately, though, direct Balmer-line observations can be used to tease out the amount of ionizations in the galaxy, and thus the amount of non-escaping ionizing photons $\xi_{\rm ion}(1-\fesc)$.
Using the \citealt{Robertson:2013bq} inference\footnote{We will assume $\fesc=0$ in all inferences of $\xi_{\rm ion}$, which conservatively underestimates the production of ionizing photons.} of $\log_{10} \xi_{\rm ion}=25.2$ Hz erg$^{-1}$ (though see~\citealt{Bouwens16xiion,Lam19xiion,barros19_xiion} for higher reported values), 
\citetalias{Robertson:2015uda} showed that $\fesc=20\%$ is sufficient if galaxies down to 0.001$L_\star$ ($M_{\rm UV}^{\rm ion.\, cutoff}\approx -13$) contribute to reionization.
We illustrate what reionization would look like for this pre-JWST calibrated model in Fig.~\ref{fig:xHIz}.
It is over by $z\sim 6$, and produces an optical depth $\tau_{\rm CMB} \approx 0.055$, bringing galaxy observations into agreement with {\it Planck} CMB measurements.

The arrival of JWST is opening a new window to reionization. 
Observations from different teams are finding large values of $\xi_{\rm ion}$, in some cases growing towards higher redshifts and fainter galaxies (though see \citealt{Endsley23_reionization} where $\xi_{\rm ion}$ is still high but grows towards the bright end instead, we study this case in Appendix~\ref{app:otherUVLF}).
In particular, \citet{Simmonds24_xiion} find a consistent increase in $\xi_{\rm ion}$ up to $z = 9$ and $\MUV=-16.5$ (where we will conservatively cap $\xi_{\rm ion}$ to avoid extrapolation), well fit by
\be
\log_{10} \left [\xi_{\rm ion}/{(\rm Hz \, erg^{-1})} \right] \approx  25.8 + 0.11  (\MUV+17) + 0.05 (z-7).
\label{eq:xi_ion}
\ee 
Such faint, early galaxies will produce $\sim4$ times more ionizing photons than expected pre-JWST~\citep[][implying a very young stellar population]{Atek24_Nature_xiion}.
The purple line in Fig.~\ref{fig:xHIz} shows how reionization would progress assuming this JWST-calibrated $\xi_{\rm ion}$, while keeping everything else the same.
In this case the additional photons would kick-start reionization by $z\sim 12$ and finish it by $z\sim 8$, far overproducing the CMB optical depth ($\tau_{\rm CMB} \approx 0.08$) when compared to observations.
Here we have kept $\fesc=0.2$ as in \citetalias{Robertson:2015uda}, so the astute reader may wonder if newer inferences of the escape fraction delay reionization.

We do not have a direct handle on $\fesc$ during reionization, as escaping ionizing photons will be absorbed by the neutral IGM before reaching us. 
However, detailed studies of low-$z$ analogues find a strong correlation, with significant scatter, between the FUV continuum slopes $\beta_{\rm UV}$ of galaxies and their LyC escape fractions. This is physically explained by the dust along the line-of-sight simultaneously attenuating the FUV stellar continuum and the ionizing photons.
We will use the fit from \citet[][calibrated on the $z\sim 0$ LzLCS survey \citealt{Flury:2022asy}, see \citealt{Trebitsch:2022jck} for an implementation on reionization]{Chisholm22_betafesc}, where
\be
\fesc = A_f \times 10^{b_f \beta_{\rm UV} }
\label{eq:fesc_Lzlcs}
\ee
with $A_f = 1.3\,\times 10^{-4}$ and $b_f = -1.22$. 
In this relation galaxies that are bluer have less dust and low-ionization gas along the line-of-sight, and thus fewer  sinks of ionizing photons. This correlation is similarly observed at $z\sim 3$ in different surveys~\citep[][]{Steidel:2018wbo,Pahl21_fesc,Begley22_VANDELSfesc,Saldana-Lopez:2022niz}.
We can then employ the $\fesc-\beta_{\rm UV}$ relation, with the $\beta_{\rm UV}-\MUV$ measurements from \citet[][which incorporates both JWST and HST measurements from~\citealt{Bouwens2014, Topping23_jwst_UVSlopes,Cullen23_jwst_UVSlopes}]{Zhao2024_betaUVJWST}
to predict\footnote{\label{foot:expavg}Note that $\VEV{\fesc}(\MUV) \equiv \int d\beta_{\rm UV} P(\beta_{\rm UV} | \MUV) \fesc(\beta_{\rm UV})$, where the PDF $P(\beta_{\rm UV} | \MUV)$ can be approximated as a Gaussian with width $\sigma_{\rm \beta}=0.34$~\citep{Smit:2012nf}, which makes $\VEV{\fesc(\beta_{\rm UV})}$ larger than $\fesc(\VEV{\beta_{\rm UV}})$ by a factor of $\exp \left[ (b_f \ln[10] \sigma_{\beta})^2 /2\right]\approx 1.1-1.5$.} $\fesc(\MUV)$.
Note that here, and throughout the text, we cap the UV slopes at $\beta_{\rm UV} = -2.7$ when computing $\fesc$ to avoid extrapolation in this relation  (though we implicitly extrapolate in $z$ and $\MUV$, see Table~\ref{tab:assumpt_summary}), as that corresponds to the bluest galaxies where Eq.~\eqref{eq:fesc_Lzlcs} is calibrated.
We show the result of applying this calibration as the blue line in Fig.~\ref{fig:xHIz}.
The JWST-calibrated $\xi_{\rm ion}$ multiplied by $\fesc$ (inferred using the high-$z$ $\beta_{\rm UV}-\MUV$ and low-$z$ $\fesc-\beta_{\rm UV}$ relations) produces an even earlier reionization, and consequently even more tension with $\tau_{\rm CMB}$.

The curves shown in Fig.~\ref{fig:xHIz} are meant to illustrate the impact of the new $\xi_{\rm ion}$ and $\fesc$ results for a particular reionization model.
Let us now move to perform a more detailed study, where we vary different underlying assumptions and compare against current observations.

\begin{figure*}
    \centering
\includegraphics[width=0.992\linewidth]{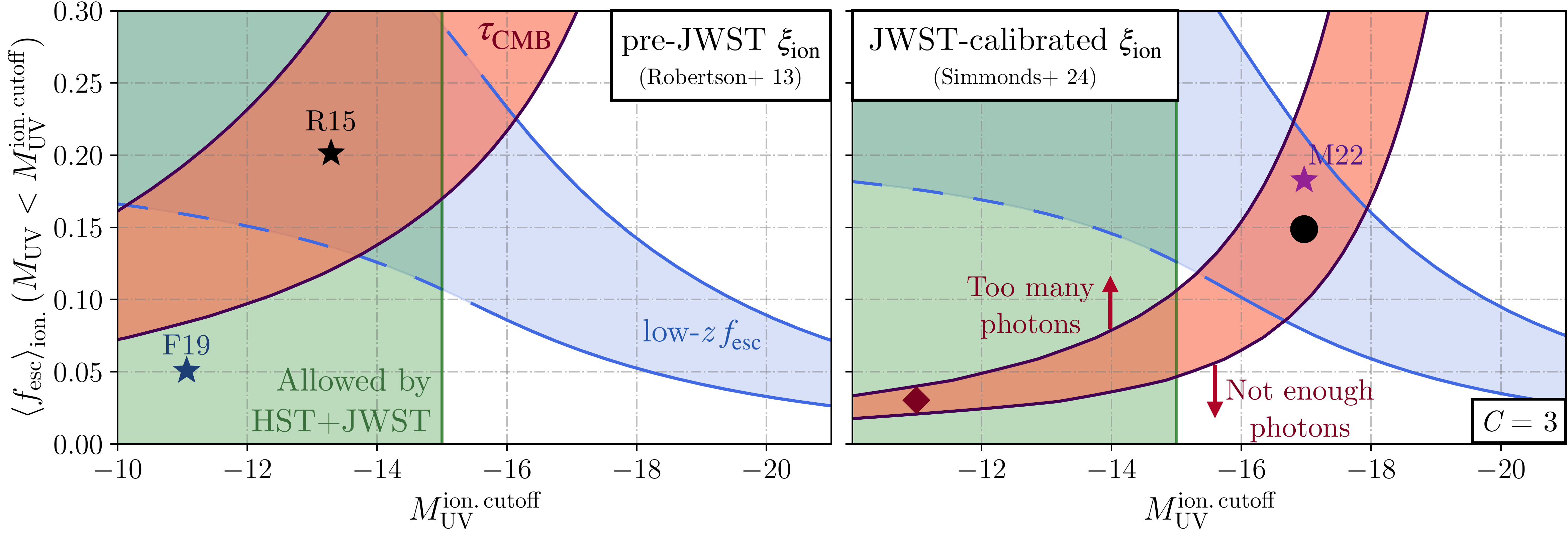}
    \caption{Tension in our models of reionization, expressed through the effective cutoff $\MUV^{\rm ion.\,cutoff}$ on the UVLF (at which galaxies cease to emit ionizing photons), and the average escape fraction $\VEV{\fesc}_{\rm ion.}$ above that cutoff.
    The three colored contours correspond to the regions allowed by the CMB optical depth $\tau_{\rm CMB}$ (red), the low-$z$ $\fesc$ studies (blue), and direct HST+JWST observations of no cutoff down to $\MUV\approx -15$ (green).
    The {\bf left} panel assumes a pre-JWST value of $\xi_{\rm ion}$ from~\citet{Robertson:2013bq}, where the three colored regions nicely overlap for faint cutoffs and $\fesc\approx 0.2$.
    The {\bf right} panel instead takes the new JWST $\xi_{\rm ion}$ calibration from~\citet{Simmonds24_xiion}, in which case {\bf the three regions do not overlap, showing a tension in reionization}.
    In more detail, the blue region follows the results from
    the LzLCS survey of reionization-era analogues~\citep[][evaluated at $z = 7$, with solid lines corresponding to the $\MUV$ directly observed, and dashed to extrapolation, in all cases capping UV slopes at $\beta_{\rm UV}=-2.7$]{Chisholm22_betafesc}.
    We highlight three popular pre-JWST models from~\citetalias{Robertson:2015uda}, \citetalias{Finkelstein:2019sbd}, and \citetalias{Matthee:2021qmg} (the latter assumes a larger $\xi_{\rm ion}$ closer to the new JWST value)
    as colored stars. 
    The red diamond and black circle on the right panel correspond to possible solutions to the tension (further explored in Fig.~\ref{fig:xHIsolutions}), which are in conflict with either the $\fesc$ or $M_{\rm UV}^{\rm ion.\, cutoff}$ constraints. 
    }
    \label{fig:fescMturn}
\end{figure*}

\vspace{-3mm}

\section{Observational Constraints}

To understand reionization we need to know the ionizing-photon budget, i.e., how many  photons are produced and what fraction escape their galaxies.
We model the former by taking the ionizing efficiency $\xi_{\rm ion}$ from Eq.~\eqref{eq:xi_ion}, as measured in JWST observations, and integrating the UVLF down to a cutoff magnitude $\MUV^{\rm ion.\, cutoff}$ (below where we will assume galaxies do not emit ionizing photons efficiently, either because $\fesc$, $\xi_{\rm ion}$, or the UVLF itself goes to zero, see Appendix~\ref{app:cutoff} for an example).
For the latter  
we define the ionization-averaged escape fraction as
\be
\VEV{\fesc}_{\rm ion.} \equiv \dfrac{\dot n_{\rm ion}(\fesc)}{\dot n_{\rm ion}(\fesc=1)},
\label{eq:fescionavg}
\ee
with $\dot n_{\rm ion}$ defined in Eq.~\eqref{eq:niondot}.
These two free parameters, $M_{\rm UV}^{\rm ion.\, cutoff}$ and $\VEV{\fesc}_{\rm ion.}$, encapsulate our uncertainty about the impact of high-$z$ galaxies on reionization. They must obey three different observational constraints.

First, the cutoff $M_{\rm UV}^{\rm ion.\, cutoff}$ has to be fainter than $-15$~\citep{Atek:2018nsc}, given current HST and JWST observations, which we show as a green band in Fig.~\ref{fig:fescMturn}. 
Second, $\VEV{\fesc}_{\rm ion.}$ should follow the constraints derived from low-$z$ analogues, shown as a blue band (using Eq.~\ref{eq:fesc_Lzlcs} with the best-fit amplitude and error from the LzLCS survey of $z\sim 0$ galaxies~\citealt{Chisholm22_betafesc}, see Fig.~\ref{fig:fescMturn_appendix} for the VANDELS $z\sim 3$ sample of \citealt{Saldana-Lopez:2022niz}).
Finally, the combination of $M_{\rm UV}^{\rm ion.\, cutoff}$ and $\VEV{\fesc}_{\rm ion.}$ have to produce the correct reionization history, which we parametrize through the CMB optical depth\footnote{We do not include measurements of $\xHI(z)$ as a constraint, but will see that the models that predict the right $\tau_{\rm CMB}$ broadly agree with them.}
\be
\tau_{\rm CMB} = \int d\ell n_e \sigma_T,
\ee
where $\ell$ is proper distance,  $\sigma_T$ is the Thomson cross section, and $n_e$ is the physical (not comoving) electron density, computed assuming that HeI reionization tracks HI, and that HeII reionization takes place at $z=4$.
The regions of parameter space that predict the correct $\tau_{\rm CMB}$ within 1$\sigma$ are shown as red bands in Fig.~\ref{fig:fescMturn}.
In the CMB bands a brighter cutoff $\MUV^{\rm ion.\, cutoff}$ requires higher values of $\VEV{\fesc}_{\rm ion.}$ to compensate the missing star formation --- and subsequent photon production --- at the faint end.

Fig.~\ref{fig:fescMturn} showcases the tension between these three observations. 
The left panel shows the pre-JWST situation, where the lower $\xi_{\rm ion}$ allowed the three observational bounds (red, blue, and green regions) to overlap over a broad swath of parameter space fainter than $\MUV^{\rm ion.\,cutoff} \approx -15$, with $\fesc\approx 15-30\%$.
The right panel, updated with the recent JWST observations, shows no overlap between the three.
In this case the requirements from $\tau_{\rm CMB}$ (red) and the low-$z$ $\fesc$ studies (blue) only overlap for cutoffs brighter than $\MUV^{\rm ion.\,cutoff} \approx -15$ (outside of the green region, and thus disfavored by direct HST+JWST observations).
In other words, the new JWST observations imply an overproduction of photons during reionization, which would end this process earlier than allowed by the CMB.
Note that the galaxies observed by JWST ($\MUV\lesssim-15$) already produce too many photons, including fainter objects would only worsen this tension.

Fig.~\ref{fig:fescMturn} also shows the parameter space of three popular reionization models:  \citetalias{Robertson:2015uda} ($\VEV{\fesc}_{\rm ion.}=0.2$ and $\MUV^{\rm ion.\, cutoff}\approx -13$), \citetalias{Finkelstein:2019sbd} (their best fit is at $\VEV{\fesc}_{\rm ion.}\approx 0.05$ and $\MUV^{\rm ion.\, cutoff}\approx -11$), and the Lyman-$\alpha$ emitter (LAE) model of \citet[][hereafter \citetalias{Matthee:2021qmg}, which we approximate as having $\VEV{\fesc}_{\rm ion.}=0.17$ for galaxies down to $\MUV^{\rm ion.\, cutoff}=-17$]{Matthee:2021qmg}.
Each of these models was calibrated to give rise to the correct
$\tau_{\rm CMB}$, though with different $\xi_{\rm ion}$ assumptions.
\citetalias{Robertson:2015uda} assumed $\log_{10} \xi_{\rm ion}\approx 25.2$ Hz erg$^{-1}$, \citetalias{Finkelstein:2019sbd} fit for a somewhat higher $z$-dependent value, whereas \citetalias{Matthee:2021qmg} used a larger $\log_{10} \xi_{\rm ion}\approx 25.8$ Hz erg$^{-1}$, calibrated to low-$z$ LAEs~\citep{Naidu:2021ryj}.
Each of these models is at odds with one of the three observational constraints, and thus outside one of the color bands in Fig.~\ref{fig:fescMturn}, either $\tau_{\rm CMB}$ \citepalias[][outside red band]{Robertson:2015uda}, $\fesc$ (\citetalias{Finkelstein:2019sbd}, blue), or $\MUV^{\rm ion.\, cutoff}$ (\citetalias{Matthee:2021qmg}, green).
As such, they illustrate three possible avenues to reduce the photon budget during reionization and reconcile galaxy and CMB observations.

\vspace{-3mm}
\section{Possible outs}

Let us now discuss possible physical mechanisms that may resolve this apparent photon budget crisis.

\begin{figure}
    \centering
    \includegraphics[width=\linewidth]{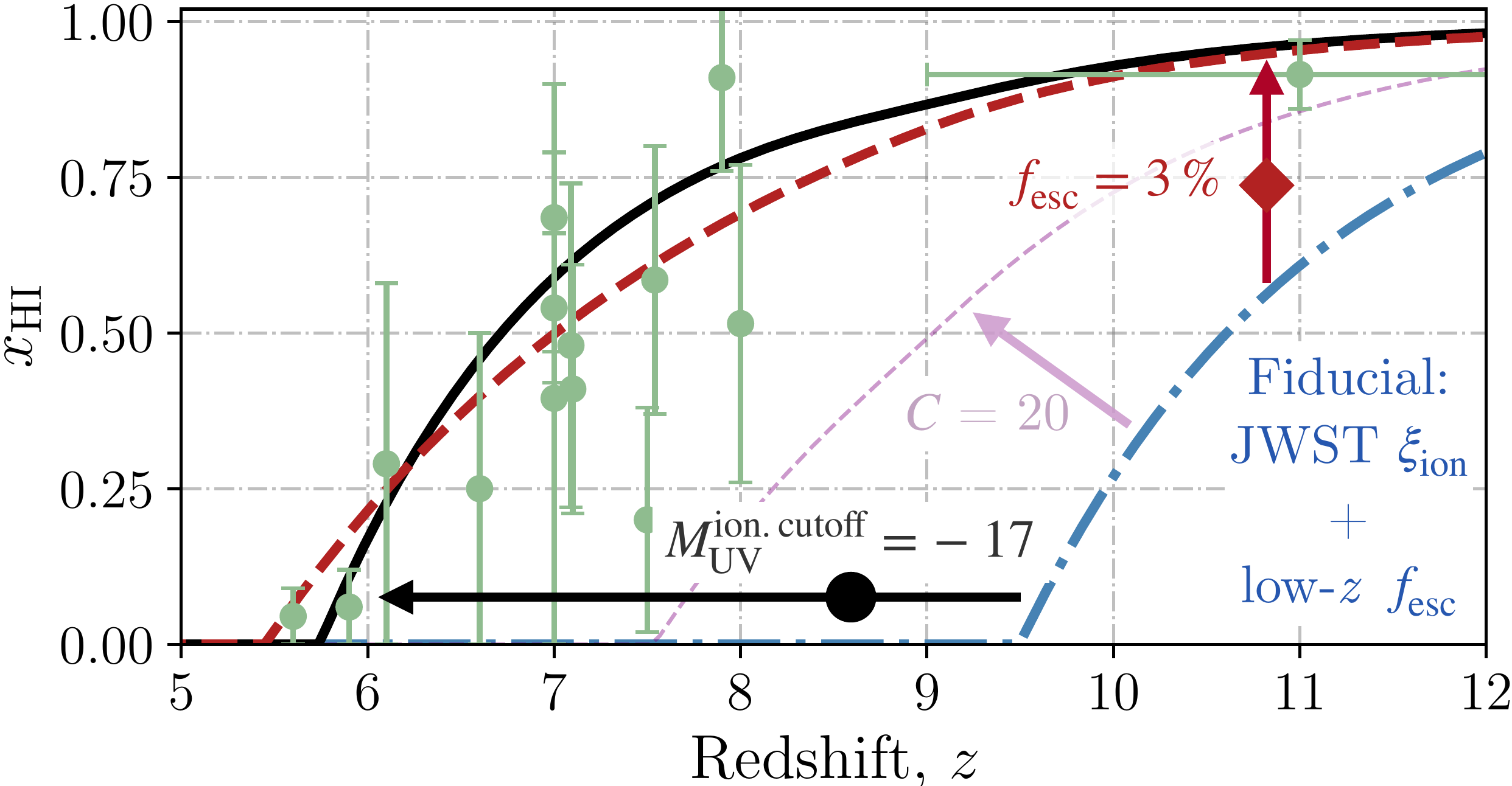}
    \caption{Possible solutions to the photon budget crisis, and how they would affect the timing of reionization. The blue dot-dashed line corresponds to our current understanding of reionization, as in Fig.~\ref{fig:xHIz}.
    The pink dotted line assumes a higher clumping factor $C=20$, which still does not produce enough recombinations.
    The red dashed line has $M_{\rm UV}^{\rm ion.\, cutoff}=-11$ and a low $\fesc=3\%$, whereas 
    the black line posits $M_{\rm UV}^{\rm ion.\, cutoff}=-17$ with a larger $\fesc=15\%$.
    These two models produce the correct $\tau_{\rm CMB}$, but disagree with one of the two galaxy observations, as indicated by the red diamond and black circle in the right panel of Fig.~\ref{fig:fescMturn}.
    Their reionization histories $\xHI(z)$ are potentially distinguishable from one another (measurements in green, which have not been used to calibrate the models).
    }
    \label{fig:xHIsolutions}
\end{figure}

$\bullet$ Perhaps some of the new $\xi_{\rm ion}$ calibrations are biased? 
It is possible that photometry alone cannot reliably recover $\xi_{\rm ion}$, that dust produces a systematic shift in this quantity~\citep{Shivaei18_dustxiion}, or that the JWST samples used to infer $\xi_{\rm ion}$ are not representative of the high-$z$ galaxy population (if they are biased towards efficient ionizers or preferentially target galaxies in a burst). For example, the sample in \citet{Simmonds24_xiion} is selected based on an emission line flux cut in photometry, which could bias the sample towards strong line emitters (and thus high $\xi_{\rm ion}$) at fixed UV magnitude.  
We have repeated our analysis with a lower fixed $\xi_{\rm ion}=10^{25.5}$ Hz erg$^{-1}$ (in line with the lowest mean values reported in \citealt{Endsley23_reionization}, which did not make an emission-line selection, as well as the $z>4$ mean in \citealt{Pahl:2024utu}, though see e.g.,~\citealt{Matthee:2022rbw} for lower values), finding that this still requires a cutoff at $\MUV \approx -14$ or brighter (see Fig.~\ref{fig:fescMturn_appendix} in Appendix~\ref{app:otherUVLF}).
An alternative solution involves keeping a high $\xi_{\rm ion}$ on average but cutting off photon production for faint galaxies (either smoothly or setting $\xi_{\rm ion}\to0$ below a cutoff magnitude $M_{\rm UV}^{\rm ion.\, cutoff}$). 
Fig.~\ref{fig:xHIsolutions} shows that a $\xi_{\rm ion}$ cutoff at $\MUV^{\rm ion.\, cutoff}=-17$ would be able to solve the tension.
Such a cutoff would, however, be in conflict with the detections of ionizing photons down to $\MUV \approx-15$ from~\citet{Atek24_Nature_xiion} and \citet[][and down to $\MUV \approx-16.5$ for the more statistically robust samples of \citealt{Simmonds24_xiion,Endsley23_reionization}]{Prieto24_xiion}.
Further JWST observations of high-$z$ galaxies will be able to determine the $\xi_{\rm ion}$ distribution down to faint magnitudes and pinpoint the impact of burstiness on this quantity.

$\bullet$ Maybe $\fesc$ is far lower than expected?
From Fig.~\ref{fig:fescMturn} it is apparent that little to no extrapolation of the LzLCS relation to bluer galaxies is required to overproduce reionization (see also Appendix~\ref{app:cutoff}).
One possibility is that the low-$z$ analogues in both LzLCS and VANDELS are biased (e.g., they may be more likely to be leakers), or that different mechanisms set $\fesc$ at high and low redshifts (so that $\fesc$ may not correlate well with $\beta_{\rm UV}$ at high $z$).  
Many of the LzLCS properties match those observed at high-redshift \citep{tang}, but it is possible (perhaps likely) that high-redshift galaxies have larger neutral gas fractions and lower dust-to-gas ratios than the low-redshift benchmarks~\citep{Heintz:2023ufb}. 
This could lead to significantly lower $\fesc$ at fixed $\beta_{\rm UV}$, or a turnaround towards fainter/bluer galaxies. While plausible, this  $\fesc-\beta_{\rm UV}$ redshift evolution is not observed in $z\sim3$ galaxies, which in fact appear to have larger $\fesc$ at fixed $\beta_{\rm UV}$ \citep{Pahl21_fesc, Saldana-Lopez:2022niz}. Another possibility is that there is a covariance between $\fesc$ and $\xi_{\rm ion}$, such that galaxies that produce large amounts of ionizing photons have lower $\fesc$. This has been predicted by simulations \citep{Rosdahl2022}, though \citet{Tang19_xiion,Naidu:2021ryj} observe the opposite trend in line emitters.

If one wanted to integrate the UVLF down to the theoretically expected cutoff at $\MUV\approx -11$~\citep{Kuhlen:2012vy}, the $\fesc$ needed to fit $\tau_{\rm CMB}$ is $\VEV{\fesc}_{\rm ion.}\approx 3$\%, as shown in Fig.~\ref{fig:xHIsolutions}, slightly lower but comparable to \citetalias{Finkelstein:2019sbd}.
For such a low value, the LzLCS relationship would require $\beta_{\rm UV} \approx -1.93$, significantly redder than JWST has observed at $z > 5$ \citep{Topping23_jwst_UVSlopes,Cullen23_jwst_UVSlopes}.
Moreover, even setting a modest $\VEV{\fesc} = 5\%$ still requires a cutoff at magnitudes brighter than $\MUV \approx -12$ given the higher $\xi_{\rm ion}$ from JWST. These faint $\MUV$ have not been statistically probed yet by JWST observations, but upcoming ultra-deep imaging of a gravitationally lensed cluster (the Glimpse program, PI: Atek, JWST ID: 3293) will measure $\xi_{\rm ion}$ and $\beta_{\rm UV}$ from lensed star-forming galaxies down to $\MUV \approx -12$.  
Deeper measurements of analogues at moderate $z$ are also critical to examine how well the $\fesc-\beta_{\rm UV}$ relation holds at fainter magnitudes (and bluer objects), as well as higher $z$, pushing as close as possible to the epoch of reionization.

$\bullet$ What about the faint end of the UVLF? The slope and turnover magnitude remain as the key uncertainties of this observable.
For a turnover to match reionization measurements it would have to be at a bright $\MUV\approx -17$, as illustrated in Fig.~\ref{fig:xHIsolutions},
far above the current UVLF limits.
An alternative is a shallow faint-end slope.
We have repeated our analysis with the \citet{FinkelsteinBagley22_UVLF} UVLF, which assumes a double power-law functional form with a flattening towards the faint end, and found that the tension persists (see Appendix~\ref{app:otherUVLF}).
This is not surprising, as the tension in Fig.~\ref{fig:fescMturn} requires little to no extrapolation of the UVLFs during reionization (down to $\MUV\approx -15$, where the faint-end uncertainties affect galaxy abundances at the $30\%$ level).
If a turnover or flattening of the UVLF was the solution it would be of paramount importance to understand its physical origin, whether it is due to feedback during reionization~\citep{Shapiro:2003gxa} or a exotic cosmology~\citep{Sabti:2021unj}.

$\bullet$ Maybe our theoretical models are wrong? The main uncertainty is how many recombinations take place, which we have modeled through a simple clumping factor $C$.
Past work has suggested additional recombinations can explain an extended reionization history inferred from the Lyman-$\alpha$ forest at $z\sim 5-6$~\citep{Davies:2021inu,Qin:2021gkn}.
Such a ``tax on the rich'' (in terms of ionizing photons, \citealt{Furlanetto:2005xx}) could alleviate the budget crisis.
As a test, we show in Fig.~\ref{fig:xHIsolutions} how even a large $C=20$ --- implying nearly an order of magnitude more recombinations throughout all of reionization --- does not suffice to harmonize galaxy and CMB observations, still overproducing reionization (more than 3$\sigma$ above $\tau_{\rm CMB}$ measurements).
Of course, we expect the process of reionization to be complex and inhomogeneous, but we note that these many recombinations per hydrogen atom are not standard in $\Lambda$CDM cosmologies~\citep[even including mini-halos,][]{Gnedin:2023cxt}, and could point to additional baryon fluctuations at very small scales, or missing ingredients in our theories.

Fig.~\ref{fig:xHIsolutions} summarizes how different possible solutions would affect the timing of reionization.
While these scenarios can be re-calibrated to produce the correct $\tau_{\rm CMB}$ (e.g., increasing $\MUV^{\rm ion.\, cutoff}$ or decreasing $\fesc$), measurements of $\xHI(z)$ can potentially distinguish between them, as posing a cutoff $\MUV^{\rm ion.\, cutoff}$ makes reionization faster (like oligarchic models,~\citealt{Naidu:2019gvi}), whereas decreasing $\fesc$ slows it down (appearing democratic,~\citetalias{Finkelstein:2019sbd}).
Of course, a $z$-dependent $\fesc$ can mimic this effect, so clustering measurements of reionization bubbles, for instance with the 21-cm line~\citep{FZH04,Munoz:2021psm} will be required to break degeneracies.
We emphasize that each of the mechanisms invoked in this section requires giving up the constraints from at least one of our galaxy measurements, be it the UVLFs, $\xi_{\rm ion}$, $\fesc$, or a combination of them.
Further observations of galaxies, $x_{\rm HI}$, and $\tau_{\rm CMB}$ will sharpen our understanding of the reionization process, as current error-bars are still sizeable and may hide underlying systematics.
While the list presented here is not exhaustive, we hope it encourages theoretical and observational work to resolve the JWST photon budget crisis.

\section{Conclusions}

The launch of JWST is allowing us to directly access the properties of the first galaxies with unprecedented sensitivity.
Early observations are showing that early, faint galaxies are prolific producers of ionizing photons.
Here we have combined new JWST measurements with determinations of the escape fractions $\fesc$ of reionization-era analogues to show that our current galaxy observations predict a process of reionization that ends too early.
That is, the situation has been reversed from the WMAP era, where the concern was producing {\it enough photons} to match $\tau_{\rm CMB}$, to the post-{\it Planck} and JWST era, where there may be {\it too many photons}.

To match the CMB optical depth, the $\dot{n}_{\rm ion}$ of early galaxies must dramatically decrease. This is currently \textit{not} observed in the high-$z$ galaxy population. For instance, the UVLFs do not show a significant turnover down to $\MUV\approx-15$~\citep{Atek:2018nsc}, faint galaxies during the epoch of reionization are very blue down to $\MUV\approx-17$  \citep[][hinting at high $\fesc$]{Topping23_jwst_UVSlopes, Cullen23_jwst_UVSlopes}, and on average galaxies produce significantly more ionizing photons than inferred from HST + \textit{Spitzer} observations down to $\MUV\approx-15$~\citep{Simmonds24_xiion, Atek24_Nature_xiion,Endsley23_reionization}. As such, taken at face value the galaxies JWST has observed already produce enough ionizing photons to reionize the universe. This does not include faint galaxies still unprobed by JWST observations, or a contribution from early black holes (which appear prevalent in early JWST observations, \citealt{Matthee:2023utn}). 
There must be a missing ingredient in either our modeling or observations to harmonize the galaxy and CMB inferences of reionization.

Moving forward, there are several avenues that can further audit the ionizing-photon budget.
Future CMB surveys are expected to measure $\tau_{\rm CMB}$ to $\approx 0.002$~\citep{LiteBIRD:2022cnt}, which would sharpen our understanding of reionization.
Better constraints on the timing of reionization beyond $\tau_{\rm CMB}$~\citep[e.g., via the kinematic Sunyaev-Zel{'}dovich effect and the transmission of Lyman-$\alpha$ photons from high-redshift sources][]{SPT-3G:2024lko,Chen2024,Nakane23_xHILAEs_JWST,Ouchi:2020zce,LuMason_bubbles24} and tomographic measurements of the distribution of neutral and ionized hydrogen through the 21-cm line~\citep{Morales:2009gs,HERA:2021noe}, will provide invaluable information on how different sources contribute to the photon budget.
Further studies of both $\fesc$ (at low $z$) and $\xi_{\rm ion}$ (at high $z$) are critical to account for selection biases in our samples, or for missed assumptions in their interpretation.
The biggest theoretical uncertainty is quantifying the recombination rate, which has a substantial effect on the reionization history (see Fig.~\ref{fig:xHIsolutions}) and may alleviate the requirements on the sources. In particular, it is crucial to understand how an increase in the recombination rate at $z \sim 6$ due to dense IGM clumps would carry over to higher redshifts, during the bulk of reionization.

In summary, recent observations have found that early galaxies were numerous, efficient producers of ionizing photons, and likely to have non-negligible escape fractions.
Together, these galaxy observations imply an excess in the ionizing-photon budget during reionization, which would end this cosmic epoch earlier than allowed by CMB data. The JWST era has just begun, and here we have examined how future observations and theoretical efforts can shed light on this tension.
As of the time of writing, the different solutions are in conflict with at least one observational constraint.
Resolving this tension on reionization is a key step to finally understanding the last major phase transition of our universe.

\section*{Acknowledgements}

We are grateful to V.~Bromm, R.~Endsley, S.~Finkelstein, K.~Hawkins, A.~Pahl, C.~Scarlata, M.~Shull, E.~Thelie, and the anonymous referee for insights on a previous version of this manuscript.
JBM was supported by the National Science Foundation under Grants AST-2307354 and AST-2408637, and thanks the Yukawa Institute for Theoretical Physics and the Kavli Institute for Theoretical Physics for their hospitality during part of this work. 
JM was supported by an appointment to the NASA Postdoctoral Program at the Jet Propulsion Laboratory / California Institute of Technology, administered by Oak Ridge Associated Universities under contract with NASA.
SRF was supported by NASA through award 80NSSC22K0818 and by the National Science Foundation through award AST-2205900. 
CAM acknowledges support by the VILLUM FONDEN under grant 37459 and the Carlsberg Foundation under grant CF22-1322. The Cosmic Dawn Center (DAWN) is funded by the Danish National Research Foundation under grant DNRF140.

\section*{Data Availability}

The data underlying this article will be shared on reasonable request to the author.
{\it Software:} numpy~\citep{numpy}, scipy~\citep{scipy}, matplotlib~\citep{matplotlib}, Zeus21~\citep{Munoz:2023kkg}, CLASS~\citep{Blas:2011rf}.



\bibliographystyle{mnras}
\bibliography{reionizationjwst} 



\appendix

\section{Alternative Assumptions}
\label{app:otherUVLF}

Through the text we have taken the pre-JWST UVLF from \citet{Bouwens21} for $z\leq 9$, and the JWST-era UVLFs from \citet{Donnan24_UVLF} for higher $z$.
These data reach $\MUV\approx -17$ at $z\sim 7$, so in order to find the abundance of galaxies at fainter magnitudes some extrapolation is required.
Additionally, we have used the $\beta_{\rm UV}-\MUV$ relation from \citet{Zhao2024_betaUVJWST}, which includes JWST data, and the $\xi_{\rm ion}$ fit from \citet{Simmonds24_xiion}. 
The purpose of this Appendix is to cross check these assumptions.
For that we will first repeat our analysis removing the new JWST calibrations of the $z\gtrsim 9$ UVLF and $\beta_{\rm UV}$ and revert to pre-JWST estimates.
Then we will use the~\citet{FinkelsteinBagley22_UVLF} UVLF, which uses a compilation of data and has a different functional form that includes flattening towards the faint end.
Finally, we will find whether there is still tension for a value of $\xi_{\rm ion}$ comparable to that of \citet{Endsley23_reionization} or \citet{Pahl:2024utu}, rather than \citet{Simmonds24_xiion}.

\begin{table*}
\begin{tabular}{l|l|l|l|l|l}
                & Relation$^a$                                                   & Instrument          & Calibrated at                                                    & Uncertainty   & $\tau_{\rm CMB}^d$          \\ \hline
$\xi_{\rm ion}$ & \citet{Simmonds24_xiion}                                   & JWST/NIRCAM         & $z\sim 4-9$, $M_{\rm UV} \lesssim -17$                           & 0.4 dex       & $0.096\pm 0.017$           \\
                & \citet{Endsley23_reionization}                             & JWST/NIRCAM         & $z\sim 6-9$, $M_{\rm UV} \lesssim -17$                           & 0.03 dex      & $0.074\pm0.001$           \\ \hline
$f_{\rm esc}-\beta_{\rm UV}$   & \citet{Chisholm22_betafesc}                                & HST/COS         & $z\sim 0$, $\beta_{\rm UV} > -2.7$ ($M_{\rm UV} \lesssim -18.5$) & 0.2 dex       & $0.096^{+0.007}_{-0.012}$ 
 \\
                & Constant 10\%                             & $-$         & $-$                           & $-$      & $0.070$   
                \\ \hline
$\beta_{\rm UV}$ & \citet{Zhao2024_betaUVJWST}                                & JWST/NIRCAM+HST/ACS & $z\sim 4-12$, $M_{\rm UV} \lesssim-16$                           & $\sim 0.2^c$ & $0.096^{+0.001}_{-0.003}$ \\
                & \citet{Bouwens2014}                                        & HST/ACS             & $z\sim 4-8$, $M_{\rm UV} \lesssim-16$                            & $\sim 0.2^c$  & $0.093^{+0.002}_{-0.004}$ \\ \hline
$\Phi_{\rm UV}$ & \citet{Bouwens21}$^{b}$                & JWST/NIRCAM+HST/ACS & $z\sim 2-14$, $M_{\rm UV} \lesssim-16$                           & $\sim 0.15^c$ dex & $0.096\pm0.007$           \\
                & \citet{FinkelsteinBagley22_UVLF}$^b$ & JWST/NIRCAM+HST/ACS & $z\sim 2-14$, $M_{\rm UV} \lesssim-16$                           & $\sim 0.15^c$ dex  & $0.076\pm0.004$          
\end{tabular}
\caption{Table summarizing the different mean/median assumed relations in this work, their origin, calibration region, and estimated uncertainty (not intrinsic scatter). Last column shows the optical depth derived by taking each relationship and its uncertainty, while keeping the rest of the analysis fixed.
$^a$ First relation shown for each variable corresponds to our fiducial through the paper. 
$^b$ In both cases added to \citet{Donnan24_UVLF} for $z\gtrsim 9$. 
$^c$ Uncertainty in $\beta_{\rm UV}$ and $\Phi_{\rm UV}$ depends on magnitude and redshift, so we report typical values at $z\sim 7$ and $\MUV\sim -17$. 
$^d$ Assuming a cutoff at $\MUV=-13$. These ought to be compared to the {\it Planck} measurement of $\tau_{\rm CMB} = 0.054\pm0.007$~\citep{Planck:2018vyg}.}
\label{tab:assumpt_summary}
\end{table*}

\begin{figure}
    \centering
\includegraphics[width=0.992\linewidth]{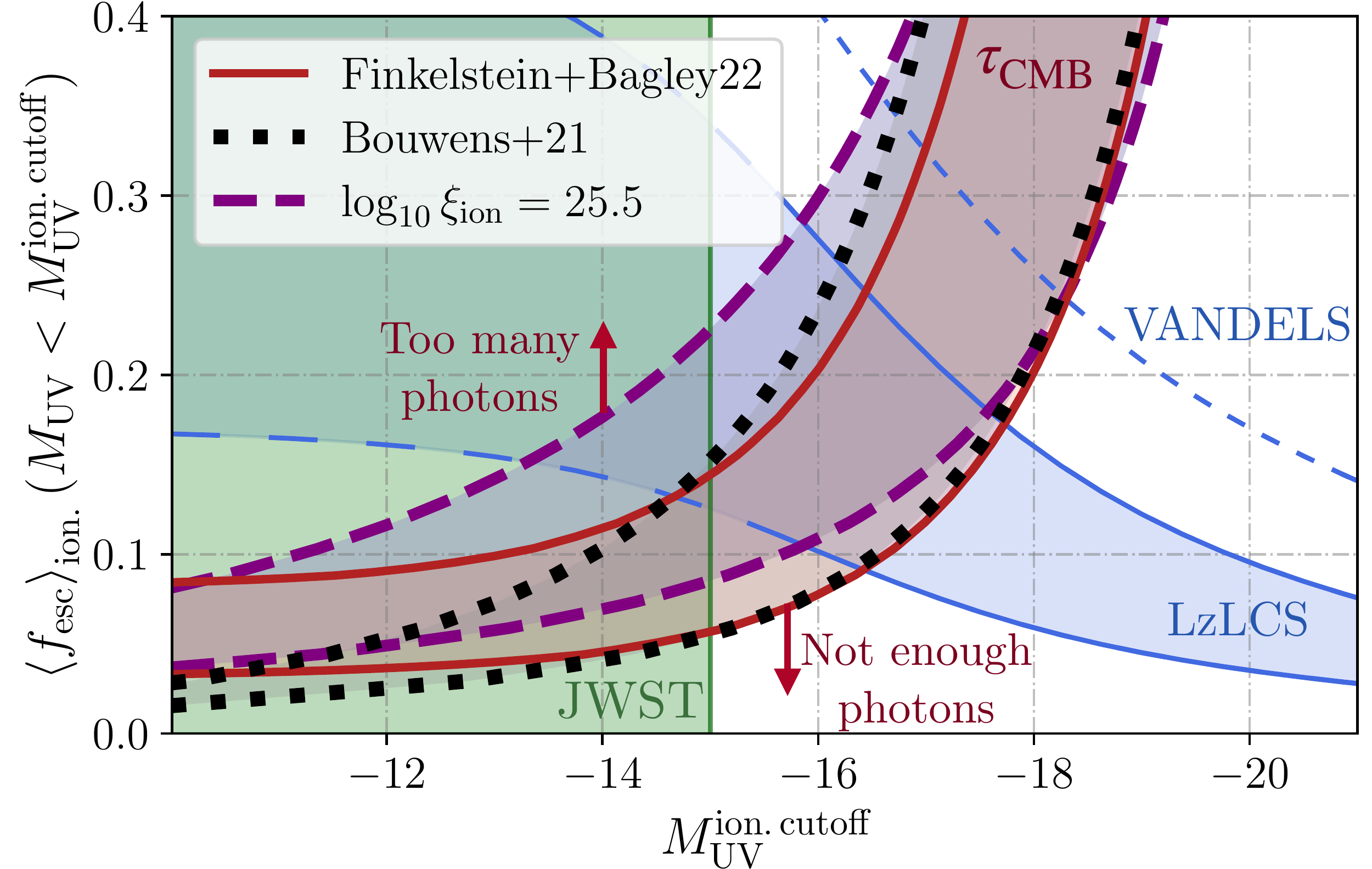}
    \caption{Constraints on reonization, as in Fig.~\ref{fig:fescMturn}, but with different assumptions about the UVLF and $\log_{10}\xi_{\rm ion}$.
    The red contour shows the $\tau_{\rm CMB}$ constraint using \citet{FinkelsteinBagley22_UVLF}, whereas black (dotted) uses that of \citet{Bouwens21}.
    Both these pre-JWST determinations of the UVLF still show a tension in reionization, though slightly less severe.
    The purple (dashed) region shows the $\tau_{\rm CMB}$ contour taking a lower value of $\xi_{\rm ion}$, which consequently predicts fewer ionizing photons.
    In this last case there is a region of parameter space where the constraints overlap, showing that a downward revision of $\xi_{\rm ion}$ plus a cutoff brighter than $\MUV\approx -14$ could resolve the tension.
    We also show the median $\fesc$ from the VANDELS sample of $z\sim 3$ galaxies~\citep{Saldana-Lopez:2022niz}, extrapolated to bluer galaxies as a dashed line.
    }
\label{fig:fescMturn_appendix}
\end{figure}

\subsection{How much do the new JWST UVLFs impact the tension?}

Not significantly. 
We re-run our analysis returning to the pre-JWST UVLF from \citet[][]{Bouwens21}, and using the $\beta_{\rm UV}-\MUV$ relation from \citet[][fixing its $z=8$ value for earlier times]{Bouwens2014}.
Fig.~\ref{fig:fescMturn_appendix} shows how the region that gives rise to the correct $\tau_{\rm CMB}$ is still in tension with galaxy observables, as it only overlaps the low-$z$ constraints on $\fesc$ for cutoffs brighter than $\MUV \approx -15$, which are disfavored (barring a tiny edge region around $\MUV^{\rm ion.\, cutoff}=-15$ and $\VEV{\fesc}_{\rm ion.}=15\%$).
The tension is then largely driven by the high $\xi_{\rm ion}$ values inferred by JWST observations, rather than the enhancement of the UVLF at high $z$.
Nevertheless, the extra $z\gtrsim 9$ galaxies can kickstart reionization earlier.
Adding the recently discovered population of supermassive black holes in JWST would potentially increase the ionizing-photon production~\citep[unlike the pre-JWST expectations, e.g.,][]{Matsuoka18_SubaruQSO_EoR}, exacerbating the crisis if the accretion disks are unobscured.

\subsection{Re-analysis with~\citet{FinkelsteinBagley22_UVLF}}

Fig.~\ref{fig:fescMturn_appendix} shows how the tension in the ionization-photon budget remains when changing the UVLF to the pre-JWST fit from \citet[][and $\beta_{\rm UV}$ from \citealt{Bouwens2014}]{FinkelsteinBagley22_UVLF}. 
The three observations ($\tau_{\rm CMB}$, the $\fesc$ measurement from low $z$, and the no-cutoff down to the HST+JWST limit) still do not overlap.
This is not surprising, since the different UVLFs broadly agree at the bright end, only diverging towards faint magnitudes and high $z$ (for instance, at $z=7$ the uncertainty in the faint-end slope $\alpha_\star$ of both \citealt{FinkelsteinBagley22_UVLF} and \citealt{Bouwens21} translates into 30\% more or fewer $\MUV=-15$ galaxies).
This is visible towards the faint side $(\MUV^{\rm ion.\, cutoff}\sim -13)$ of Fig.~\ref{fig:fescMturn_appendix}, where the  CMB-preferred region flattens at $\fesc\approx 6\%$, whereas in the \citet{Bouwens21} case it does so at $\fesc \approx 3\%$. 
Part of the reason is the turnover built into the UVLF fit of \citet[][not included in the \citealt{Bouwens21} fit]{FinkelsteinBagley22_UVLF}, regardless of our additional $\MUV^{\rm ion.\, cutoff}$.
This test serves to benchmark the differences in the faint end of the UVLF.

\subsection{A lower $\xi_{\rm ion}$ value?}

Through the main text we have used the fit for $\xi_{\rm ion}$ as a function of $\MUV$ and $z$ from \citet{Simmonds24_xiion}.
Other reionization-era results from  \citet{Atek24_Nature_xiion} and \citet{Endsley23_reionization} also find enhanced ionizing-photon production, though in the latter case it decreases towards the faint end, rather than increase.
We have found that the photometric results in \citet{Endsley23_reionization} can be approximately fit by
\be
\log_{10} \xi_{\rm ion} = 25.5 - 0.03 \,\times (\MUV+18),
\ee
for the two faint bins in their calibration (and this relation underestimates $\xi_{\rm ion}$ for the brightest bin).
We show in Table~\ref{tab:assumpt_summary} how taking this relation still overpredicts $\tau_{\rm CMB}$.
A recent spectroscopic analysis in \citet{Pahl:2024utu} finds a mean $\VEV{\log_{10} \xi_{\rm ion}/(\rm Hz\, erg^{-1})} = 25.38$ for their $z>4$ sample, which translates into a mean $\VEV{\xi_{\rm ion}} = 10^{25.57} \rm Hz\, erg^{-1}$ (Pahl, Private Communication) as expected of a lognormal variable with a 0.4 dex scatter.
We can, then, conservatively bracket the uncertainty in $\xi_{\rm ion}$ by performing a run with $\log_{10}\xi_{\rm ion}/(\rm Hz\, erg^{-1}) = 25.5$, comparable to the lowest mean values measured in \citet[][see third panel of Fig.~\ref{fig:nioncutoff}]{Endsley23_reionization} and the running mean of \citet{Pahl:2024utu}.
We show the result of this analysis in Fig.~\ref{fig:fescMturn_appendix}, where the three observational constraints overlap over a small range of parameter space.
This represents a possible compromise solution, requiring both a cutoff brighter than $\MUV\sim -14$ (potentially detectable) plus a downward revision on $\xi_{\rm ion}$ (possibly indicating an observational bias or mismodeling).

\subsection{Summary of Assumptions}

We summarize the relations taken in this work, their origin in either JWST or HST data, and the reported uncertainties in Table~\ref{tab:assumpt_summary}.
We compute in each case the expected $\tau_{\rm CMB}$ and associated errorbars from each relationship, keeping the rest fixed and setting a fiducial cutoff at $\MUV=-13$ as in \citet{Robertson:2015uda}.
We find that changing the UVLF calibration makes a difference of $20\%$ on $\tau_{\rm CMB}$ (due to the cutoff included in \citealt{FinkelsteinBagley22_UVLF}), whereas changing the $\beta_{\rm UV}$ relation is at the sub-10\% level. 
The biggest uncertainties are $\fesc$ and $\xi_{\rm ion}$, as expected, and in particular we find that going from the \citet{Simmonds24_xiion} to the \citet{Endsley23_reionization} $\xi_{\rm ion}$ calibration  reduces $\tau_{\rm CMB}$ by 20\%, though in all cases shown in Table~\ref{tab:assumpt_summary} $\tau_{\rm CMB}$ is higher than allowed by the CMB.

\section{The origin of a faint cutoff}
\label{app:cutoff}

\begin{figure}
    \centering
\includegraphics[width=0.8\linewidth]{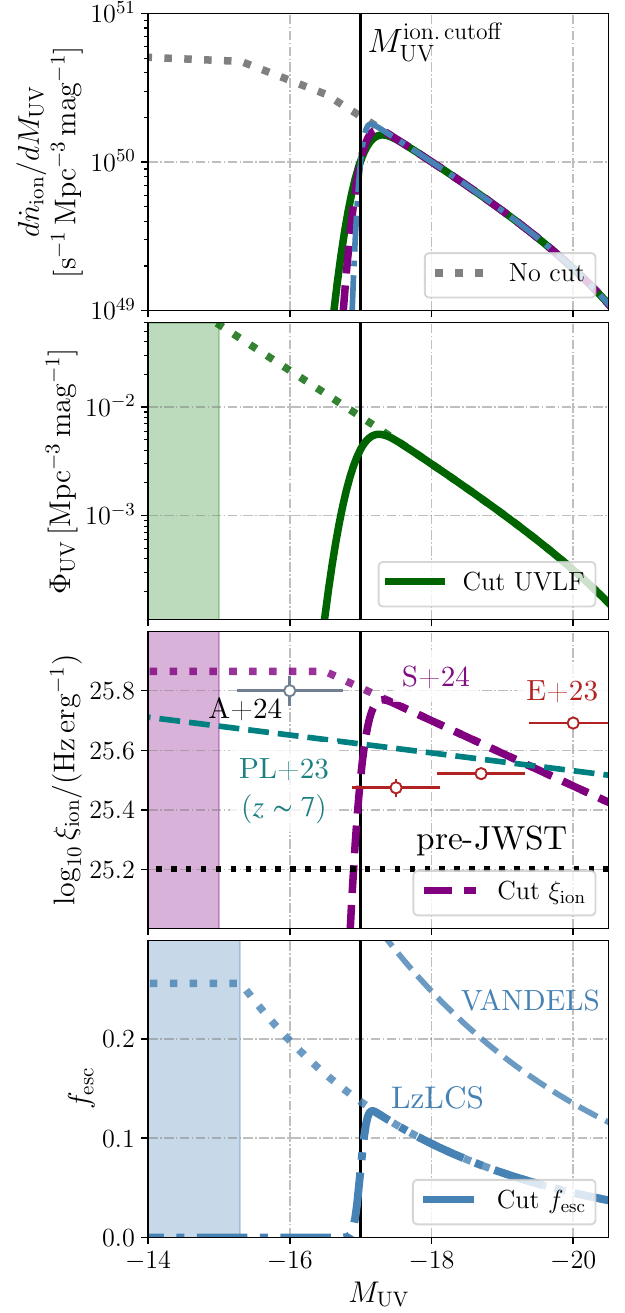}
    \caption{The ionization-photon production ({\bf first panel}, at $z\sim 7$) can be cut off at the faint end from three different sources: the UVLF ($\Phi_{\rm UV}$, {\bf second panel}), the ionizing efficiency ($\xi_{\rm ion}$, {\bf third panel}), or the escape fraction ($\fesc$, {\bf last panel}).
    We shade the regions that are {\bf not} observed on each respective panel.
    Dotted curves show the result with no cutoffs, and thick with each cutoff. The vertical black line is at $\MUV^{\rm ion.\, cutoff}=-17$, as required to fit reionization in Fig.~\ref{fig:xHIsolutions}, which is in tension with observations.
    In the third panel we show not only the $\xi_{\rm ion}$ fit from \citet[][S+24, used through the main text]{Simmonds24_xiion} but also measurements from \citet[][E+23]{Endsley23_reionization}, \citet[][A+24]{Atek24_Nature_xiion}, and \citet[][PL+23, evaluated at $z\sim 7$]{Prieto24_xiion}. All the JWST-inferred $\xi_{\rm ion}$ values are well above the pre-JWST canonical value (black dotted).
    }
    \label{fig:nioncutoff}
\end{figure} 

Through the main text we have used the variable $\MUV^{\rm ion.\, cutoff}$ to express a generic cutoff below where galaxies do not contribute to reionization.
This cutoff can have an origin in three different mechanisms, which we illustrate in Fig.~\ref{fig:nioncutoff}.

First, the UVLF may have a ``turn over'', so the abundance of star-forming galaxies drops below some magnitude.
Current UVLF observations suggest that this cutoff has to be fainter than $\MUV\approx-15$ during reionization~\citep{Atek:2018nsc}. 
This is shown in the second panel of Fig.~\ref{fig:nioncutoff}.

Second, the ionizing efficiency may vanish for faint galaxies.
Some JWST observations suggest the opposite, in fact, with $\xi_{\rm ion}$ seemingly growing towards the faint end at least until $\MUV\approx-15$~\citep[][or $\MUV=-16.5$ for the broader but photometric sample of~\citealt{Simmonds24_xiion}]{Atek24_Nature_xiion}.
This trend is also reported in \citet{Prieto24_xiion} at $\VEV{z} \approx 4$, which we show in Fig.~\ref{fig:nioncutoff} (extrapolating their results to $z = 7$ by using the scaling in Eq.~\ref{eq:xi_ion}).
The results in \citet{Endsley23_reionization} instead point to $\xi_{\rm ion}$ growing towards the bright end, as shown in Fig.~\ref{fig:nioncutoff}, though with a large variance in the distribution at each bin.
This variance also makes the average $\xi_{\rm ion}$ slightly larger than expected from the median of $\log_{10}\xi_{\rm ion}$, as in footnote \ref{foot:expavg}, which we account for when plotting the \citet{Endsley23_reionization} data (as it is the only one with measured variance).
All the JWST measurements in the third panel of Fig.~\ref{fig:nioncutoff} are significantly above the pre-JWST canonical value, following pre-JWST hints in e.g.,~\citet{Maseda2020}.

Last, the escape fraction $\fesc$ may stop growing towards the faint end.
This is, however, the opposite behavior seen at low $z$ in both LzLCS ($z\sim 0$) and VANDELS ($z\sim 3$).
For reference, the median $\fesc$ in the VANDELS sample reported by \citet{Saldana-Lopez:2022niz} can be fit using Eq.~\eqref{eq:fesc_Lzlcs} with $A_f = 1.12 \times 10^{-4}$ and $b_f = -1$ (from their Fig.~15), which we show  in both Figs.~\ref{fig:fescMturn_appendix} and \ref{fig:nioncutoff}.
The bluest galaxies sampled in the LzLCS ($\beta_{\rm UV}=-2.7$) correspond to $\MUV\approx -15.7$ at $z\sim 7$ (using the $\beta_{\rm UV}-\MUV$ relation from~\citealt{Zhao2024_betaUVJWST}), indicating we do not expect a cutoff on $\fesc$ until at least that magnitude, as shown in the last panel of Fig.~\ref{fig:nioncutoff}.

Together, different galaxy observations have probed the $\MUV^{\rm ion.\, cutoff}\approx-17$ region, disallowing a cutoff at such magnitudes, unless there is an observational bias or systematic uncertainty.

\bsp	
\label{lastpage}
\end{document}